\documentclass[sigconf]{acmart}

\usepackage{amsmath,amsfonts}
\usepackage{algorithmic}
\usepackage{textcomp}
\usepackage{xcolor}
\usepackage{booktabs}
\usepackage{array}
\usepackage{multirow}
\usepackage{multicol}
\usepackage{enumitem}
\usepackage{subfig}
\usepackage{bm}
\usepackage{balance}
 
\usepackage{amssymb}

\newcommand{\Mat}[1]{\bm{#1}}
\newcommand{\Vector}[1]{\bm{#1}}

\newcommand{\Loss}{\mathcal{L}}
\newcommand{\Normal}[1]{\mathrm{#1}}

\newcommand{\QSet}{\mathcal{Q}}   
\newcommand{\ISet}{\mathcal{I}}   

\newcommand{\FC}{\mathrm{FC}}

\newcommand{\Softmax}{\mathrm{Softmax}}

\AtBeginDocument{%
  \providecommand\BibTeX{{%
    \normalfont B\kern-0.5em{\scshape i\kern-0.25em b}\kern-0.8em\TeX}}}

\copyrightyear{2023}
\acmYear{2023}
\setcopyright{acmlicensed}
\acmConference[CIKM '23]{Proceedings of the 32nd ACM International Conference on Information and Knowledge Management}{October 21--25, 2023}{Birmingham, United Kingdom.}
\acmBooktitle{Proceedings of the 32nd ACM International Conference on Information and Knowledge Management (CIKM '23), October 21--25, 2023, Birmingham, United Kingdom}
\acmPrice{15.00}
\acmDOI{10.1145/3583780.3615457}
\acmISBN{979-8-4007-0124-5/23/10}

\begin{document}

\title{Beyond Semantics: Learning a Behavior Augmented Relevance Model with Self-supervised Learning}

\author{Zeyuan Chen}
\affiliation{
  \institution{Ant Group}
  \country{}
}
\email{chenzeyuan.czy@antgroup.com}

\author{Wei Chen}
\affiliation{
  \institution{Ant Group}
  \country{}
}
\email{qianmu.cw@antgroup.com}

\author{Jia Xu$^{*}$}
\thanks{*Corresponding author.}
\affiliation{
  \institution{Ant Group}
  \country{}
}
\email{steve.xuj@antgroup.com}

\author{Zhongyi Liu}
\affiliation{
  \institution{Ant Group}
  \country{}
}
\email{zhongyi.lzy@antgroup.com}

\author{Wei Zhang}
\affiliation{
  \institution{East China Normal University}
  \country{}
}
\email{zhangwei.thu2011@gmail.com}

% \renewcommand{\shortauthors}{Chen et al.}
\renewcommand{\shortauthors}{Zeyuan Chen, Wei Chen, Jia Xu, Zhongyi Liu, \& Wei Zhang}

\begin{abstract}
Relevance modeling aims to locate desirable items for corresponding queries, which is crucial for search engines to ensure user experience. Although most conventional approaches address this problem by assessing the semantic similarity between the query and item, pure semantic matching is not everything. In reality, auxiliary query-item interactions extracted from user historical behavior data of the search log could provide hints to reveal users' search intents further. Drawing inspiration from this, we devise a novel Behavior Augmented Relevance Learning model for Alipay Search (BARL-ASe) that leverages neighbor queries of target item and neighbor items of target query to complement target query-item semantic matching.
Specifically, our model builds multi-level co-attention for distilling coarse-grained and fine-grained semantic representations from both neighbor and target views. The model subsequently employs neighbor-target self-supervised learning to improve the accuracy and robustness of BARL-ASe by strengthening representation and logit learning. Furthermore, we discuss how to deal with the long-tail query-item matching of the mini apps search scenario of Alipay practically. Experiments on real-world industry data and online A/B testing demonstrate our proposal achieves promising performance with low latency.
\end{abstract}




\maketitle

\section{Introduction}\label{sec:intro}
In this era of information overload, search engines are indispensable for Internet content platforms, enabling users to efficiently locate desirable items among the multitude of candidates for corresponding queries. To guarantee the experience for users, relevance modeling is essential to maintain an adequate level of relevance between query and presented results, which serves as a fundamental component of search engines. 

In the relevant literature, early works~\cite{robertson2009probabilistic,shah2010evaluating,svore2009machine} perform feature engineering to finish text-based matching without enough generalization and accuracy. Then deep learning-based approaches emerge as the new paradigm, which can be broadly classified into \textit{representation-based} approaches~\citep{shen2014learning,palangi2014semantic,rao2019bridging} and \textit{interaction-based} approaches~\citep{parikh2016decomposable,chen2016enhanced,hu2014convolutional,pang2016text}. In recent times, pre-trained models like BERT~\citep{devlin2018bert} have made remarkable strides in Natural Language Understanding (NLU) tasks. As a result, a few studies~\cite{yao2022reprbert,lu2020twinbert,reimers2019sentence} are proposed to encode semantic correlations within queries and items. Most recently, Large Language Models (LLMs) have demonstrated their superior ability in various downstream Natural Language Processing (NLP) tasks. These models, such as GPT~\citep{radford2019language}, LLaMA~\citep{touvron2023llama}, and GLM~\citep{du2022glm}, are trained on massive corpora of texts, which enables them to learn the underlying patterns and structures of natural language. However, it remains difficult to accurately identify user search intents solely based on semantics. The texts of queries and items in Alipay search are quite short and ambiguous, making it hard to convey effective information contained in their identity. For example, given a query ``Zhe Yi'', the abbreviation of a hospital, it is challenging to comprehend the actual semantics. But its historical clicked items include ``the first affiliated hospital of Zhejiang University'', indicating strong correlations between them to help search intent understanding. As such, leveraging behavior data to assist relevance modeling is a natural strategy. Moreover, combining textual and user behavior data can lead to comprehensive searches, as it allows for a more holistic understanding of users' search intents.

Existing studies~\cite{yao2021learning,zeng2022graph,zhu2021textgnn,li2021adsgnn,pang2022improving} have primarily conducted the use of user behavior data but have not fully explored the potential insights that can be gained from behavioral data. They all consider constructing the pre-training dataset or the graph structure based on click behaviors. Nevertheless, the former inevitably introduces noises into the ground truth without human annotation and further leads to the confusion of relevance modeling. And the click graph may amplify the effect of noises through the multi-hop neighborhood aggregation scheme, making the learning more vulnerable to interaction noises. This may account for the sub-optimal performance of TextGNN involving high-order neighbors in A/B testing shown in Section~\ref{sec:ab}. Besides, Graph Neural Networks (GNNs)~\cite{hamilton2017inductive,velivckovic2017graph,zhang2021graph,chen2021learning} have been observed to learn representations that heavily rely on nodes and edges, overlooking the potential correlations among neighbor nodes. It restricts the ability of the GNNs to fully capture the complex relationships and patterns within a graph. But fine-grained interaction bridging the correlations of the all inputs explicitly and fully is essential for semantic matching tasks~\cite{yang2016anmm,zhang2021graph,yin2016abcnn}.

To overcome the drawbacks of existing methods, we propose the BARL-ASe model that formulates historical query-item interactions as dual behavior neighbors for target query and item. In contrast to behavior neighbors (queries) of a candidate item that reflect the semantic preference, behavior neighbors (items) of a given query reveal its real intent. Specifically, we perform fine-grained interaction on dual behavior neighbors and target query/item with multi-level co-attention to comprehensively explore the local and global information. Then we anticipate reducing the discrepancy between query (item) behavior neighbors and target item (query) which should embody similar semantic information ideally and mitigating the impact of noises through contrastive learning to obtain abundant and robust representations. In order to take advantage of the information from neighbor and target views to compensate for each other, we employ mutual learning to achieve view-level alignment, which can be regarded as a specialized version of contrastive learning without uniformity. Finally, we propose an industrial application for online serving, which combines the strengths of neighbor-dependent/independent models in Alipay search to maximise proceeds. 

In summary, we make the following contributions of this paper:
\begin{itemize}
    \item To the best of our knowledge, we are the first to effectively leverage dual behavior neighbors for corresponding queries and items to enable the adaptable combination of fine-grained interactions and topology information.
    
    \item We propose BARL-ASe with two main components.
    Firstly, a multi-level co-attention is developed to learn fine-grained interactions from the perspective of the target and neighbor.
    Secondly, the proposal of self-supervised learning can strengthen the fusion and alignment of neighbor and target view, which in turn improves the model performance.

    \item We explore an industrial application to deal with long-tail query-item matching through the combination of neighbor-dependent/independent models efficiently and effectively powered by LLMs.
    
    \item We demonstrate BARL-ASe achieves superior performance through experiments on real-world industry data. It has been deployed online and outperforms prior approaches on core metrics.
\end{itemize}

\section{Related Work}\label{sec:related}
The section concretely discusses the most recent advances in relevance modeling studies. Relevance modeling in search can be viewed as a text matching problem as the sub-domain of information retrieval (IR). Current approaches can be classified into two aspects: \textit{feature-based} approaches and \textit{deep learning-based} approaches.
The first category is centered on manual-crafted features such as TF-IDF similarity and BM25~\citep{svore2009machine}. Despite their usefulness, these feature-based approaches have limited generalization ability due to their domain-specific features and require significant labor resources. 

In order to address the limitations of the above approaches, deep learning-based approaches emerge as the new paradigm, which can be broadly classified into \textit{representation-based} approaches and \textit{interaction-based} approaches. The former focuses on learning a low-dimensional representation of data while the latter emphasizes capturing the interaction between inputs. For instance, DSSM~\citep{shen2014learning} is a classical two-tower representation-based model that encodes the query and the document separately. In this paradigm, recurrent~\citep{palangi2014semantic,tai2015improved} and convolutional~\citep{hu2014convolutional,shen2014learning} networks are adopted to extract low-dimensional semantic representations. For these methods, the encoding of each input is carried out independently of the others. Consequently, these models face challenges in modeling complex relationships. To overcome this limitation, interaction-based models are proposed. DecompAtt~\citep{parikh2016decomposable} leverages attention network to align and aggregate representations. In parallel, recurrent~\citep{chen2016enhanced} and convolutional~\citep{hu2014convolutional,pang2016text} networks are employed for modeling complex interactions.

In recent times, pre-trained models like BERT~\citep{devlin2018bert} have made remarkable strides in Natural Language Understanding (NLU). As a result, representation-based~\citep{yao2022reprbert,lu2020twinbert,reimers2019sentence} and interaction-based architectures~\citep{wang2019structbert} are proposed to leverage the capabilities of these models to encode semantic correlations between queries and items. Most recently, Large Languages Models (LLMs) like GPT~\citep{radford2019language}, LLaMA~\citep{touvron2023llama}, GLM~\citep{du2022glm} trained on massive corpora of texts have shown their superior ability in language understanding, generation, interaction, and
reasoning tasks. ~\citep{sun2023chatgpt} investigates the potential of utilizing LLMs for searching and demonstrates that appropriately instructs ChatGPT and GPT-4 can produce competitive and even superior results to supervised methods widely used information retrieval benchmarks. In this work, we try to explore the applications of LLMs in relevance modeling stage of Alipay search engine.

In addition to textual information, there are related works that aim to integrate user behavior data into their models. The utilization of user behavior data can provide valuable insights into search intent, which can then be used to enhance the relevance of the search engines. MASM~\citep{yao2021learning} leverages the historical behavior data to complete model pre-training as a weak-supervision signal with a newly proposed training objective. ~\citep{zhu2021textgnn,li2021adsgnn,pang2022improving} try the incorporation of click graphs to enhance the effectiveness of search systems. In contrast to existing works that center on learning the high-order topology structure while disregarding the fine-grained interaction among behavior neighbors, our work endeavors to explore thoroughly the natural implementation of adaptable semantic textual information fusion with behavior neighbors while considering interaction granularity and topology structure. Our research seeks to highlight the importance of interaction granularity and topology structure in search engines, and its potential to improve the performance of such systems in real-world scenarios.

\section{Problem Formulation}\label{sec:problem}
Assume we have the target query $\Mat{q}$ and target item $\Mat{i}$ needed to predict the relevance degree, where $\Mat{q} = (\Vector{w}^q_1,\Vector{w}^q_2,...,\Vector{w}^q_{l(q)})$ and $\Mat{i} = (\Vector{w}^i_1,\Vector{w}^i_2,...,\Vector{w}^i_{l(i)})$ denote a composition of $l(q)$ and $l(i)$ word embedding respectively. Besides, we use $\QSet^i$ and $\ISet^q$ to denote behavior neighbors of the target item $\Mat{i}$ and target query $\Mat{q}$, which are obtained based on a query-item interaction graph constructed from past click behaviors (Over a past month). To mitigate the effect of noises, query-item interaction pairs are selected above a click-through threshold (e.g., 0.2) and the neighbors are arranged in descending order based on click-through rate. The behavior neighbors $\QSet^i$ of target item $\Mat{i}$ with length of $l(\QSet)$ are denoted as $\QSet^i = (\Vector{q}^i_1,\Vector{q}^i_2,...,\Vector{q}^i_{l(\QSet)})$, where $\Vector{q}^i_{l}$ represents the $l$-th neighbor query embedding for target item $\Mat{i}$. Similarly, the behavior neighbors $\ISet^q$ corresponding to target query $\Mat{q}$ are denoted $\ISet^q = (\Vector{i}^q_1,\Vector{i}^q_2,...,\Vector{i}^q_{l(\ISet)})$. The neighbor embeddings are obtained by performing mean-pooling operations over contained word embeddings. Given the above notations, the aim of relevance modeling is to learn a function: $f(\Mat{q}, \Mat{i}, \QSet^i, \ISet^q) \rightarrow \hat{y}_{qi}$ to generate score $\hat{y}_{qi}$ representing the relevance degree with the inputs of target query, target item, and behavior neighbors.

\begin{figure*}[!t]
    \centering
    \includegraphics[width=.75\linewidth]{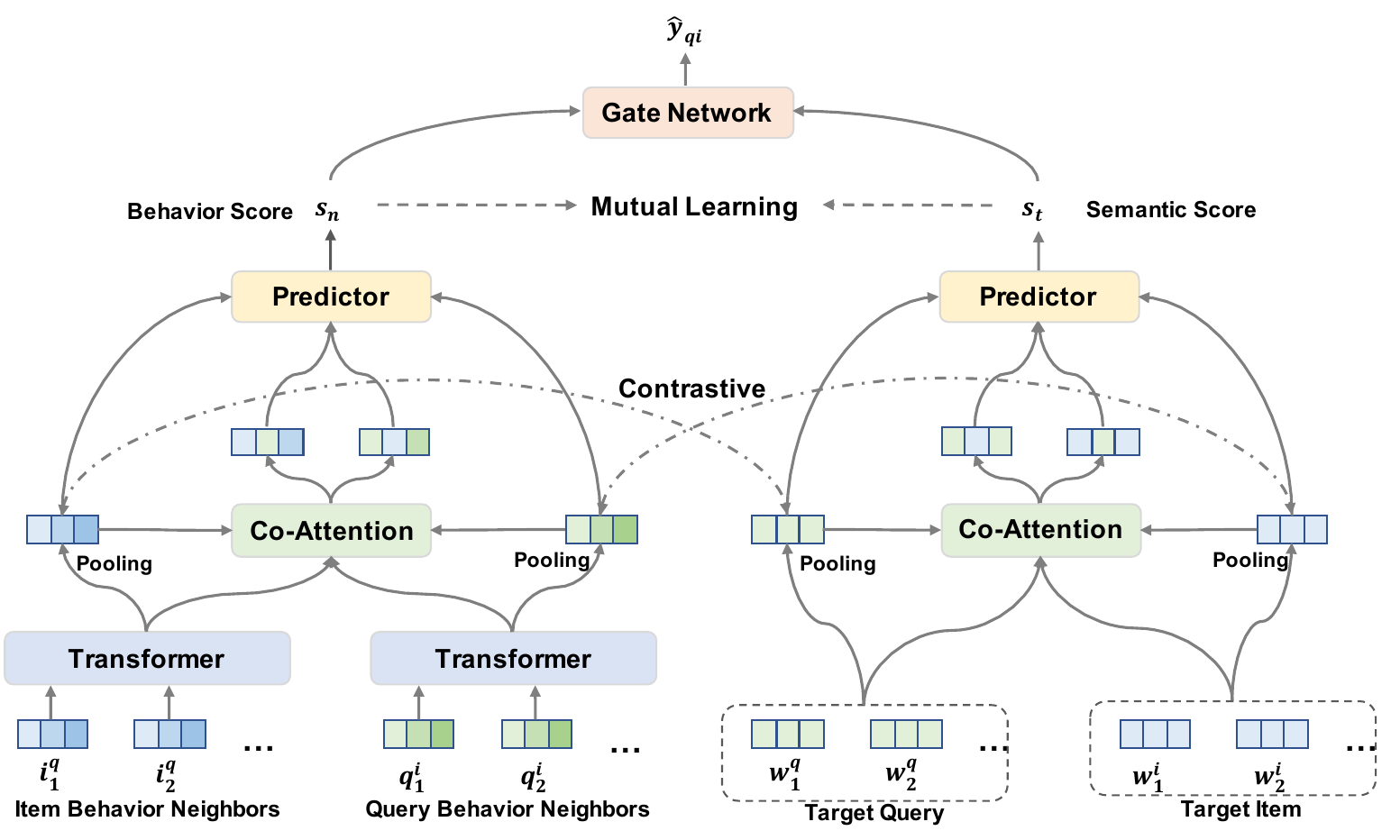}
    % \vspace{-1.em}
    \caption{The architecture of BARL-ASe.}
    \label{fig:architecture}
    % \vspace{-1.em}
\end{figure*}

\section{Methodology}\label{sec:method}
The architecture of BARL-ASe is depicted in Figure~\ref{fig:architecture}. It contains
two fundamental modules: (1) multi-level co-attention for learning fine-grained neighbors and target representations; (2) neighbor-target self-supervised learning for improved representations. Based on the representations obtained by the two modules, relevance score generation can be performed. In what follows, we elaborate on the two modules, followed by the generation of relevance scores and the training stage. 

\subsection{Multi-level Co-attention}
Before delving into the introduction of multi-level co-attention, we first clarify how to encode the input embeddings. Given the powerful capabilities demonstrated by Transformers~\cite{vaswani2017attention} in various areas~\cite{kang2018self,wu2022self,devlin2018bert}, it is logical to employ them to model the inputs.
Encoding the behavior neighbors except words of the target yields improved performance with a customized transformer which shares the parameters of self-attention module and performs a feed-forward network of varied parameters w.r.t. different behavior neighbors~\cite{wang2022image}. For the sake of simplicity, we shall adhere to the notations used above after providing inputs to the transformer. We first perform a pooling operation over dual behavior neighbors for obtaining the coarse-grained global representations as follows:
\begin{equation}\label{eq:behavior-pooling}
    \Vector{h}^{\QSet}_p = \Normal{Pooling}(\QSet^i)\,,\quad\quad \Vector{h}^{\ISet}_p = \Normal{Pooling}(\ISet^q)\,,
\end{equation}
where the pooling operation can be chosen as mean-pooling, sum-pooling, self-attention, etc. Here we choose mean-pooling based on the local experiments. Similarly, we can obtain the representations of the target denoted as $\Vector{h}^{q}_p$ and $\Vector{h}^{i}_p$. Then we employ neighbor-level co-attention for dual behavior neighbors to comprehensively explore the local and global information, which can be formulated as:
\begin{equation}\label{eq:co-att}
    \Vector{h}^{\QSet}_a, \Vector{h}^{\ISet}_a = \Normal{Co-Att}(\Vector{h}^{\QSet}_p,\Vector{h}^{\ISet}_p,\QSet^i,\ISet^q) = \sum_{l=1}^{l(\ISet)}\alpha_l \Vector{i}^{q}_{l},\sum_{l=1}^{l(\QSet)}\beta_l \Vector{q}^{i}_{l},
\end{equation}
\begin{equation}
    \alpha_l = \Softmax(\omega^\QSet\tanh(\Mat{W}^\QSet[\Vector{h}^{\QSet}_p\oplus \Vector{i}_l^q]+\Vector{b}^\QSet)),
\end{equation}
\begin{equation}
    \beta_l = \Softmax(\omega^\ISet \tanh(\Mat{W}^\ISet[\Vector{h}^{\ISet}_p\oplus \Vector{q}_l^i]+\Vector{b}^\ISet)),
\end{equation}
where $\alpha_l,\beta_l$ are attention weights. The trainable parameters are $\{\omega^\QSet, \Mat{W}^\QSet, \Vector{b}^\QSet, \omega^\ISet, \Mat{W}^\ISet, \Vector{b}^\ISet\}$. Analogously, we use the following manner, $\Vector{h}^{q}_a, \Vector{h}^{i}_a =\Normal{Co-Att}(\Vector{h}^{q}_p,\Vector{h}^{i}_p,\Mat{q},\Mat{i})$, to represent target-level co-attention for target query and item.

\subsection{Neighbor-target Self-supervised Learning}
\subsubsection{Neighbor-target Contrastive Learning}
$\Vector{h}^{q}_p$ and $\Vector{h}^{i}_p$ denote the actual semantics of the target query and item. And $\Vector{h}^{\QSet}_p$ and $\Vector{h}^{\ISet}_p$ represent the indirect representations reflecting the potential intent of the target query and item. We anticipate enhancing performance from a dual view through the fusion and alignment of actual semantics and potential intent. As such, we follow SimCLR~\cite{chen2020simple} and adopt InfoNCE to maximize the agreement of positive pairs and minimize that of negative pairs:
\begin{equation}\label{eq:infonce}
\Loss(\Vector{h}^{\QSet}_p,\Vector{h}^i_p) = -\log\frac{\exp(g(\Vector{h}^{\QSet}_p,\Vector{h}^i_p)/\tau)}{\sum_{k}^K \exp(g(\Vector{h}^{\QSet}_p,\Vector{h}^i_{pk})/\tau)},
\end{equation}
where $g(\cdot)$ is a similarity function between two vectors (e.g., cosine similarity) and $\tau$ is a temperature hyper-parameter. $\Vector{h}^i_{pk}$ denotes the $k$-th negative sample dissimilar with $\Vector{h}^{\QSet}_p$ of total $K$ negative samples in a batch. Likewise, we acquire the contrastive loss of target query $\Loss(\Vector{h}^{\ISet}_p,\Vector{h}^q_p)$. The overall loss can be denoted as $\Loss_c = \Loss(\Vector{h}^{\QSet}_p,\Vector{h}^i_p)+\Loss(\Vector{h}^{\ISet}_p,\Vector{h}^q_p)$.

\subsubsection{Neighbor-target Mutual Learning} Apart from the aforementioned neighbor-target contrastive learning, we perform mutual learning between dual views with the perspective of logits, formulated as follows:
\begin{equation}
    \Loss_m = (s_n-s_t)^2\,,
\end{equation}
\begin{equation}\label{eq:behavior-pooling}
    s_n=\FC(\Vector{n})\,,\quad\quad s_t = \FC(\Vector{t})\,, 
\end{equation}
\begin{equation}
    \Vector{n} = [\Vector{h}^{\QSet}_p\oplus\Vector{h}^{\QSet}_a\oplus\Vector{h}^{\ISet}_p\oplus\Vector{h}^{\ISet}_a]\,,\quad\quad \Vector{t}=[\Vector{h}^{q}_p\oplus\Vector{h}^{q}_a\oplus\Vector{h}^{i}_p\oplus\Vector{h}^{i}_a]\,, 
\end{equation}
where $\Vector{n}$ and $\Vector{t}$ represent the concatenation of the representations from the neighbor and target view, respectively. $s_n$ denotes the behavior score computed from the neighbor view and the semantic score generated based on semantic texts is denoted as $s_t$. The logits from dual views could be well exploited to compensate each other with $\Loss_m$, thereby mitigating model bias. $\FC$ denotes the fully-connected layer.

\subsection{Model Prediction and Training}\label{sec:train}
 For the relevance scores $s_n$ and $s_t$ of dual views, we can estimate their contributions via a gating mechanism. We define the following formula to compute the relative importance weight $\gamma$:
 \begin{equation}
 \gamma = \sigma(\omega[\Vector{n}\oplus \Vector{t}])\,, 
 \end{equation}
 where $\sigma$ is the sigmoid function, and $\omega$ is a trainable parameter. Then we obtain the final relevance score for the target query and item by the weighted summation of $s_n$ and $s_t$:
 \begin{equation}\label{eq:main_loss}
     \hat{y}_{qi} = \gamma \cdot s_n + (1-\gamma) \cdot s_t\,, 
 \end{equation}
 For training the model BARL-ASe, we minimize the cross-entropy loss of the predicted relevance score w.r.t. the ground truth. The loss function is given by:
 \begin{equation}
     \Loss_{main} = -\left(y_{qi}\,\Normal{log}\,\hat{y}_{qi}+(1-y_{qi})\,\Normal{log}\,(1-\hat{y}_{qi})\right)\,, 
 \end{equation}
 In the end, we unify the main relevance modeling task and the self-supervised learning task for jointly optimizing the model. The hybrid objective function can be formulated as:
 \begin{equation}\label{eq:loss}
     \Loss = \Loss_{main}+\lambda_1\Loss_c+\lambda_2\Loss_m\,, 
 \end{equation}
 where $\lambda_1$ and $\lambda_2$ are hyper-parameters to control the strengths of self-supervised learning.

 \section{Long-tail Query-Item Pairs Matching}
 The effectiveness of the BARL-ASe model in enhancing relevance matching performance is contingent on the inclusion of behavior neighbors. Consequently, behavior neighbors play a critical role in our proposed approach. However, long-tail query-item pairs associated with only a limited number or even none of the query/item neighbors may potentially impact the performance of our proposals as shown in figure~\ref{fig:com}. Therefore, the consideration of such factors is crucial in the evaluation and optimization of our approach. In order to achieve optimal performance regardless of the presence of behavior neighbors, we propose the model denoted as BARL-ASe$_+$, which combines neighbor-dependent/independent models efficiently and effectively. BARL-ASe is a neighbor-dependent model obviously and we choose one of LLMs as our neighbor-independent model owing to their superior zero-shot or few-shot abilities. Specifically, we employ our proposed BARL-ASe model to generate relevance scores for those query-item pairs associated with query/item neighbors and Alipay-customized large-scale language model (i.e., AntLLM) pre-trained on massive corpora of texts of Alipay to process with long-tail query-item pairs. AntLLM (0.1B) fine-tuned by our human-annotated dataset is deployed online considering resources and efficiency. Due to the policy of Alipay, we avoid discussing the details of AntLLM. Here we introduce how to apply AntLLM to relevance modeling in our scenario. Referring to PET~\citep{schick2020exploiting}, we formulate the pattern $P$ based on the inputs, which is then processed by AntLLM $M$ to determine which verbalizer $v$ is the most likely substitute for the mask. For our task, the pattern can be formulated as follows:
 \begin{equation}
     P(\Mat{q},\Mat{i})=Is\,[\Mat{q}]\,related\,to\,[\Mat{i}]?\,[mask].
 \end{equation}
 The verbalizer could be "No" or "Yes" by mapping the label $y_{qi}\in\{0,1\}$ to a word from the vocabulary of AntLLM to denote the relevance degree between $\Mat{q}$ and $\Mat{i}$. Utilizing the logits obtained from $M(v(y_{qi})\,|\,P(\Mat{q},\Mat{i}))$, we could obtain the final relevance score through $\Softmax$ function.

\section{Experiments}\label{sec:exp}
\subsection{Experimental Setup}
\subsubsection{Datasets}
In order to evaluate the performance of all the models with reliability, we select the real-world industry data used in mini apps search scenario of Alipay search engine and present its statistics in Table~\ref{tbl:stat}. The dataset is labeled by human annotators, where Good and Bad annotations denote label 1 and 0 respectively. User historical behavior data is sampled from the search log of the vertical search engine. Although datasets such as WANDS\footnote{\url{https://github.com/wayfair/WANDS/tree/main}} and MSLR\footnote{\url{https://www.microsoft.com/en-us/research/project/mslr/}} are publicly available, they do not contain the requisite user historical behavior data. Hence, we select this in-house data to evaluate the proposed model, which is larger than the public datasets available. 

\subsubsection{Baseline Models}
\begin{itemize}[leftmargin=*]
    \item \textbf{DSSM}~\cite{shen2014learning} is a two-tower representation-based model. It encodes the embedding of a given query and item independently and computes the relevance score accordingly.
    
    \item \textbf{ReprBert}~\cite{yao2022reprbert} is a representation-based Bert model that utilizes novel interaction strategies to achieve a balance between representation interactions and model latency.
    
    \item \textbf{Bert}~\cite{devlin2018bert} has achieved significant progress on NLP tasks as an interaction-based model. Here we concatenate the query and item as the input of the model.
    
    \item \textbf{MASM}~\cite{yao2021learning} leverages the historical behavior data to complete model pre-training as a weak-supervision signal with a newly proposed training objective.
    
    \item \textbf{TextGNN}~\cite{zhu2021textgnn} extends the two-tower model with the complementary graph information from user historical behaviors.
    
    \item \textbf{AdsGNN}~\cite{li2021adsgnn} further proposes three aggregation methods for the user behavior graph from different perspectives.

    \item \textbf{AntLLM} is a powerful LLM with different magnitude of parameters customized by Alipay. Despite models with 2B or larger parameters are currently not supported in our scenario due to substantial resource requirements and inference latency, we still conduct an evaluation to verify the effectiveness of BARL-ASe.
\end{itemize}

\begin{table}[!t]
\centering
\caption{Statistics of the human-annotated dataset.}\label{tbl:stat}
\vspace{-1.em}
\begin{tabular}{cccccc}
\hline
Dataset  & \# Sample  & \# Query  & \# Item  & \# Good & \# Bad\\ \hline
 Train & 773,744 & 87,499 & 88,724 & 460,610 & 313,134\\
 Valid & 97,032 & 40,754 & 25,192 & 57,914 & 39,118\\
 Test & 96,437 & 40,618 & 25,004 & 57,323 & 39,114\\
\hline
\end{tabular}
\vspace{-1.em}
\end{table}

\subsubsection{Evaluation Metrics}
We use Area Under Curve (AUC), F1-score (F1), and False Negative Rate (FNR) to measure the performance of models. AUC and F1 are commonly used in the studied area, of which higher metric values represent better model performance. Conversely, lower False Negative Rate (FNR) values are preferable, as they indicate a lower false filtering rate of models. Note that AUC often serves as the most significant metric in our task while the others provide auxiliary supports for our analysis.

\subsubsection{Model Implementations}
We optimize our proposed model BARL-ASe by Adam with a learning rate of 0.0005 and a mini-batch size of 1024. The dimension of the representations is 256. The number of behavior neighbors is set to 20 and the layer of the transformer is set to 1. We add L2 regularization to the loss function (Equation~\ref{eq:loss}) by setting the weight to 1e-4. We tune three $\lambda_1,\lambda_2$ and $\tau$ within the ranges of \{0.01, 0.05, 0.1, 0.2, 0.5, 1.0\}, \{0.01, 0.05, 0.1, 0.2, 0.5, 1.0\}, and \{0.1, 0.2, 0.5, 1.0\}, respectively. $\FC$ layer is set to 2 for non-linear transformation with layer sizes 256 and 128. The implementation and partial data of this paper are available\footnote{\url{https://github.com/alipay/BehaviorAugmentedRelevanceModel}}.

\begin{table}[!t]
\centering
\caption{Main results on real-world industry data. The best and second-best performed methods in each metric are highlighted in bold and underlined, respectively. (-) denotes the lower value corresponds to better performance. Improvements over variants are statistically significant with p < 0.05.}\label{tbl:performance-comp}
\vspace{-0.5em}
\resizebox{0.8\linewidth}{!}{
\begin{tabular}{c|ccc} % |*{7}{c|}
\hline
Method
&\multicolumn{1}{c}{AUC} &\multicolumn{1}{c}{F1} &\multicolumn{1}{c}{FNR (-)}  \\ \hline
\hline

DSSM
&\multicolumn{1}{c}{0.8356} &\multicolumn{1}{c}{0.8210} &\multicolumn{1}{c}{0.1389}  \\

ReprBert
&\multicolumn{1}{c}{0.8388} &\multicolumn{1}{c}{0.8376} &\multicolumn{1}{c}{0.1280} \\

Bert
&\multicolumn{1}{c}{0.8540} &\multicolumn{1}{c}{0.8534} &\multicolumn{1}{c}{0.1150} \\

MASM
&\multicolumn{1}{c}{0.8547} &\multicolumn{1}{c}{0.8318} &\multicolumn{1}{c}{0.1289} \\

TextGNN
&\multicolumn{1}{c}{0.8847} &\multicolumn{1}{c}{0.8489} &\multicolumn{1}{c}{0.1290} \\

AdsGNN
&\multicolumn{1}{c}{\underline{0.8878}} &\multicolumn{1}{c}{0.8458} &\multicolumn{1}{c}{0.1454} \\

AntLLM (0.1B)
&\multicolumn{1}{c}{0.8597} &\multicolumn{1}{c}{0.8567} &\multicolumn{1}{c}{0.1145} \\

AntLLM (2B)
&\multicolumn{1}{c}{0.8619} &\multicolumn{1}{c}{0.8598} &\multicolumn{1}{c}{0.1065} \\

AntLLM (10B)
&\multicolumn{1}{c}{0.8751} &\multicolumn{1}{c}{\underline{0.8656}} &\multicolumn{1}{c}{\textbf{0.1006}} \\

\hline
\hline
BARL-ASe
&\multicolumn{1}{c}{0.9078} &\multicolumn{1}{c}{0.8658} &\multicolumn{1}{c}{0.1054}  \\
BARL-ASe$_+$
&\multicolumn{1}{c}{\textbf{0.9096}} &\multicolumn{1}{c}{\textbf{0.8675}} &\multicolumn{1}{c}{\underline{0.1021}}  \\
\hline
\end{tabular}
}
\vspace{-0.6em}
\end{table}

\begin{figure}[!h]
    \centering
    \includegraphics[width=0.7\linewidth]{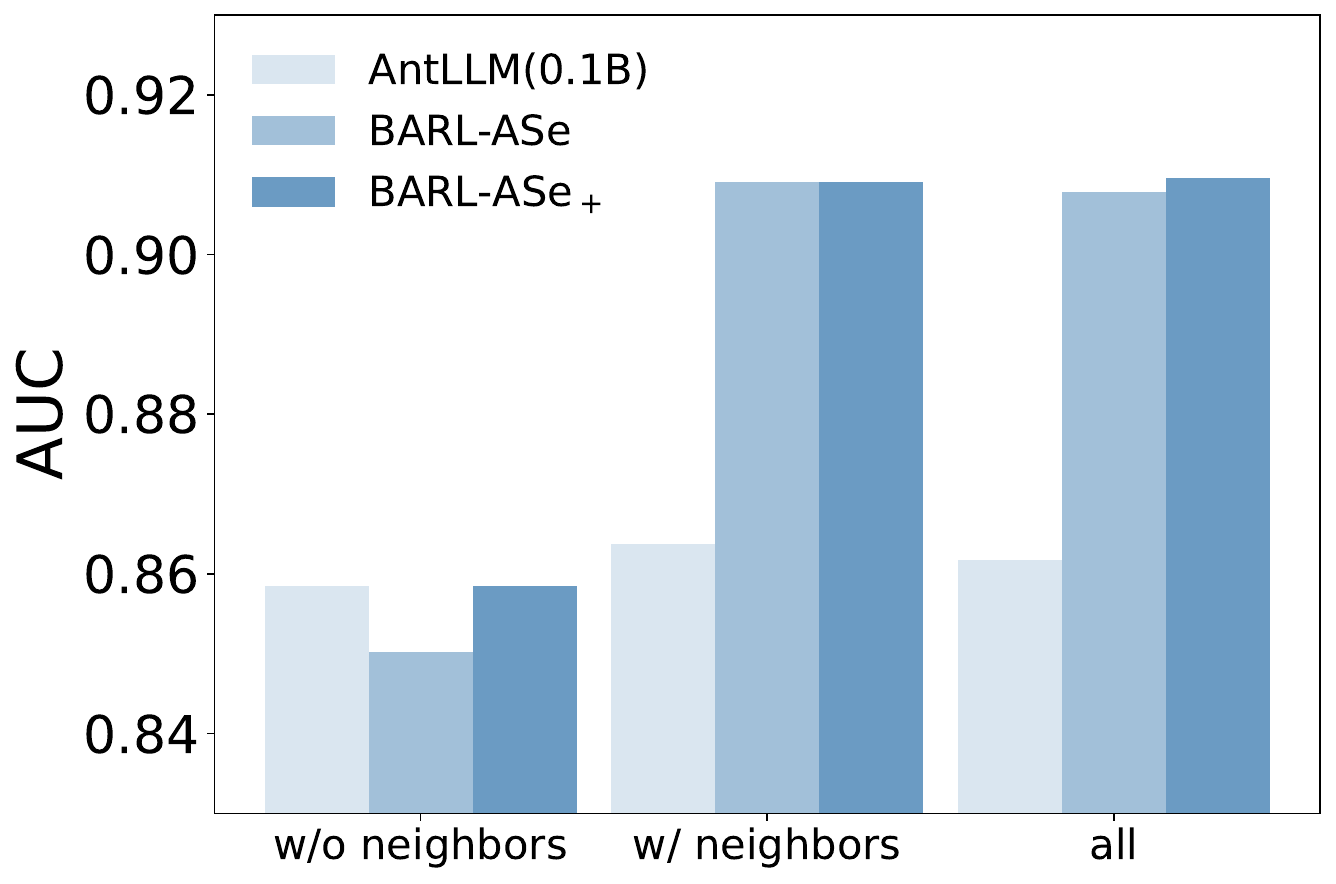}
    % \vspace{-1.em}
    \caption{Performance comparison of deployed models on neighbor-aware dataset.}
    \label{fig:com}
    \vspace{-1.em}
\end{figure}

\subsection{Experimental Results}
\subsubsection{Performance Comparison}
Table~\ref{tbl:performance-comp} shows the overall comparison of BARL-ASe and BARL-ASe$_+$ with different baselines.
Our observations show that DSSM performs poorly, which is in line with expectations, as it only encodes the embeddings of query and item separately as a basic two-tower model. By further comparing DSSM with ReprBert, we find the performance is improved to a certain degree. This demonstrates the pre-trained models utilizing the large corpus can bring additional gains. Compared to ReprBert, Bert achieves better performance. This is understandable, as the interaction-based models possess more advanced text relevance modeling abilities than representation-based models relying on query and item representations solely. 

For the models of MASM, TextGNN, and AdsGNN, they exploit historical behavior data as an auxiliary signal and achieve significantly better results than the above-mentioned methods in the metric of AUC rather than F1 and FNR. This can be attributed to the fact that the introduction of auxiliary signals. It may bring improvement for the ranking ability of models but introduce some noises leading to the loss of F1 and FNR. The performance differences among them depend on the utilization of behavior data. 

AntLLM exhibits relatively good performance from a whole perspective, demonstrating the positive effect of massive training datasets and large-scale parameters. Among
them, models with larger-scale parameters achieve better performance. Large-scale parameters may improve the capacities of models studied from massive training datasets. Compared to our model, AntLLM (10B) achieves comparative performance in F1 and FNR. However, the requirement of substantial GPU resources and delays of hundreds of milliseconds of AntLLM (2B or larger parameters) are not acceptable for industrial scenarios from an economic and efficiency point of view. BARL-ASe only requires CPU resources with an average latency of only about 13 milliseconds.

Besides, we divide the test dataset into three categories (neighbor-aware dataset) to verify the rationality of our approach. "w/o neighbors" denotes the dataset containing the query-item pairs associated with no neighbors. Similarly, "w/ neighbors" contains neighbors for query-item pairs. "all" represents all test dataset. As shown in figure~\ref{fig:com}, the neighbor-independent model (AntLLM) proceeds similar performance among these datasets. And the resulting performance difference is large for neighbor-dependent model (BARL-ASe) in neighbor-aware dataset. Thus, BARL-ASe$_+$ is proposed to efficiently maximize the performance. However, only 6.3\% of the data has no neighbors, which may account for minor improvements over the test dataset. 

Overall, our model BARL-ASe$_+$ yields the best performance and ensures efficiency and economy with about 0.2B parameters.
To be specific, BARL-ASe$_+$ obtains 2.46\% gains under AUC and achieves the best F1 and FNR on the evaluation dataset compared to models that could be deployed in our scenario.

\begin{table}[!h]
\centering
\vspace{-0.6em}
\caption{Ablation study of BARL-ASe}\label{tbl:abl}
\vspace{-0.6em}
\resizebox{0.7\linewidth}{!}{
\begin{tabular}{c|ccc} % |*{7}{c|}
\hline
Method
&\multicolumn{1}{c}{AUC} &\multicolumn{1}{c}{F1} &\multicolumn{1}{c}{FNR (-)}  \\ \hline
\hline

w/o QBN
&\multicolumn{1}{c}{0.9005} &\multicolumn{1}{c}{0.8612} &\multicolumn{1}{c}{0.1128}  \\

w/o IBN
&\multicolumn{1}{c}{0.8827} &\multicolumn{1}{c}{0.8467} &\multicolumn{1}{c}{0.1212} \\
\hline
w/o NCA
&\multicolumn{1}{c}{0.9005} &\multicolumn{1}{c}{0.8540} &\multicolumn{1}{c}{0.1219} \\

w/o TCA
&\multicolumn{1}{c}{0.8967} &\multicolumn{1}{c}{0.8489} &\multicolumn{1}{c}{0.1246} \\
\hline
w/o QNTC
&\multicolumn{1}{c}{0.9051} &\multicolumn{1}{c}{0.8608} &\multicolumn{1}{c}{0.1167} \\

w/o INTC
&\multicolumn{1}{c}{0.9038} &\multicolumn{1}{c}{0.8578} &\multicolumn{1}{c}{0.1297} \\
\hline
w/o Mutual
&\multicolumn{1}{c}{0.9032} &\multicolumn{1}{c}{0.8598} &\multicolumn{1}{c}{0.1220} \\
\hline
\hline
BARL-ASe
&\multicolumn{1}{c}{\textbf{0.9078}} &\multicolumn{1}{c}{\textbf{0.8658}} &\multicolumn{1}{c}{\textbf{0.1054}}  \\
\hline
\end{tabular}
}
\vspace{-1.em}
\end{table}

\subsubsection{Ablation Study}
To investigate the contributions of key components and types of behavior neighbors adopted by BARL-ASe, we provide the following variants of our complete model:
(1) ``w/o QBN'' denotes discarding the input of query behavior neighbors for the target item;
(2) similarly, ``w/o IBN'' denotes discarding the input of item behavior neighbors;
(3) ``w/o NCA'' keeps the self-supervised learning part but removes the neighbor-level co-attention as shown in Equation~\ref{eq:co-att};
(4) ``w/o TCA'' represents removing the target-level co-attention network;
(5) ``w/o QNTC'' denotes removing the neighbor-target contrastive learning of target query shown in Equation~\ref{eq:infonce};
(6) analogously, ``w/o INTC'' discards the neighbor-target contrastive learning of target item;
(7) ``w/o Mutual'' means not considering neighbor-target mutual learning.

Throughout the ablation study shown in Table~\ref{tbl:abl}, we observe that:
\begin{itemize}[leftmargin=*]
    \item By analyzing the results of ``w/o QBN'' and ``w/o IBN'', we could conclude that learning from dual behavior neighbors is profitable. And the type of item behavior neighbors is more significant than the other type.
    
    \item ``w/o NCA'' and ``w/o TCA'' show the comparable performance of the multi-level co-attention network, where TCA is more effective than NCA. It is likely that target-level co-attention attempts to identify the direct semantic correlations between the target query and item.
    
    \item The comparison between ``w/o QNTC'', ``w/o INTC'', and the full model validates differentiating the contrastive learning of target query and item can consistently boost the relevance modeling task.
    
    \item ``w/o Mutual'' is inferior to the full model BARL-ASe, verifying that using neighbor-target mutual learning is advantageous for deriving an effective information ensemble.
\end{itemize}

\subsection{Parameter Sensitivity}
\begin{figure}[!h]
    \centering
    \vspace{-1.em}
    \subfloat
    {
    \includegraphics[width=0.31\linewidth]{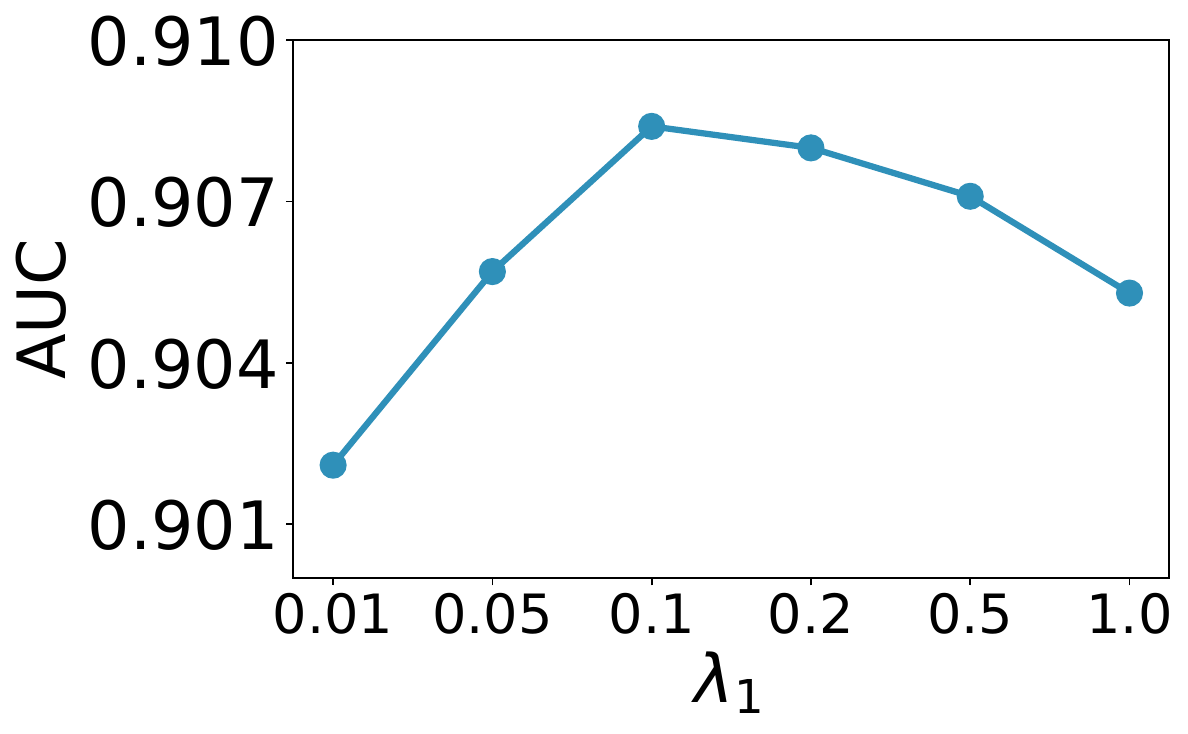}\label{fig:lambda1}
    \includegraphics[width=0.31\linewidth]{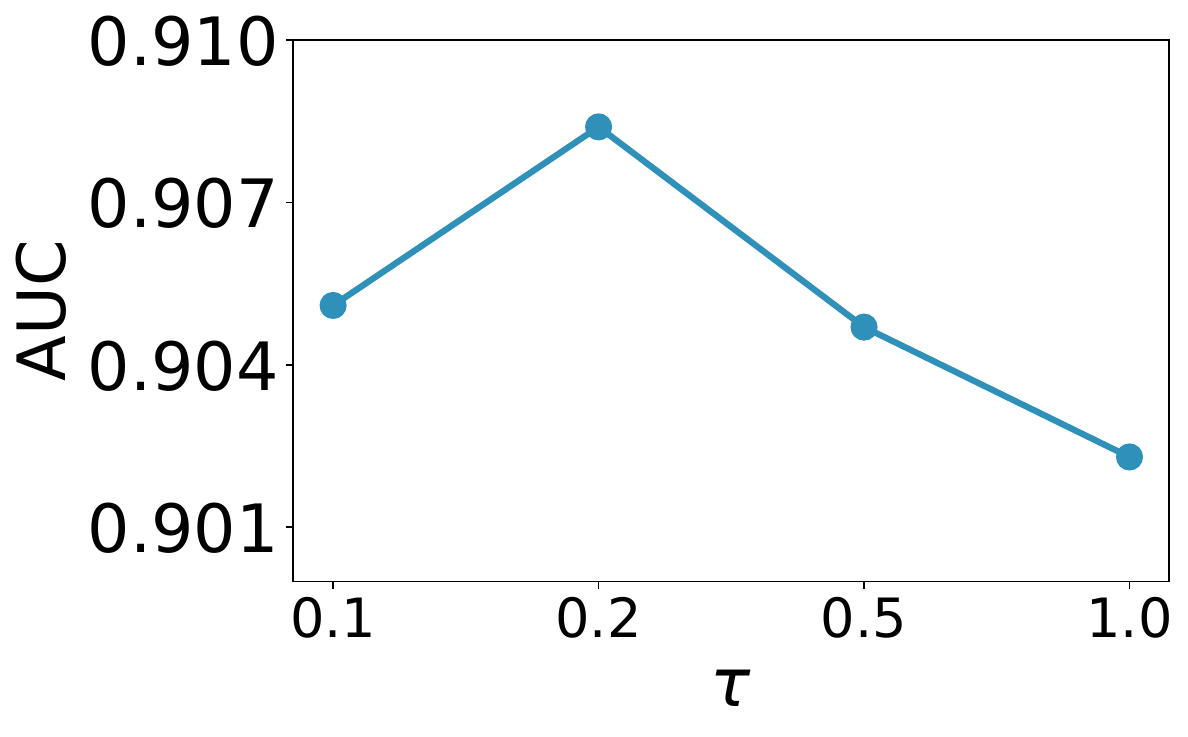}\label{fig:tau}
    \includegraphics[width=0.31\linewidth]{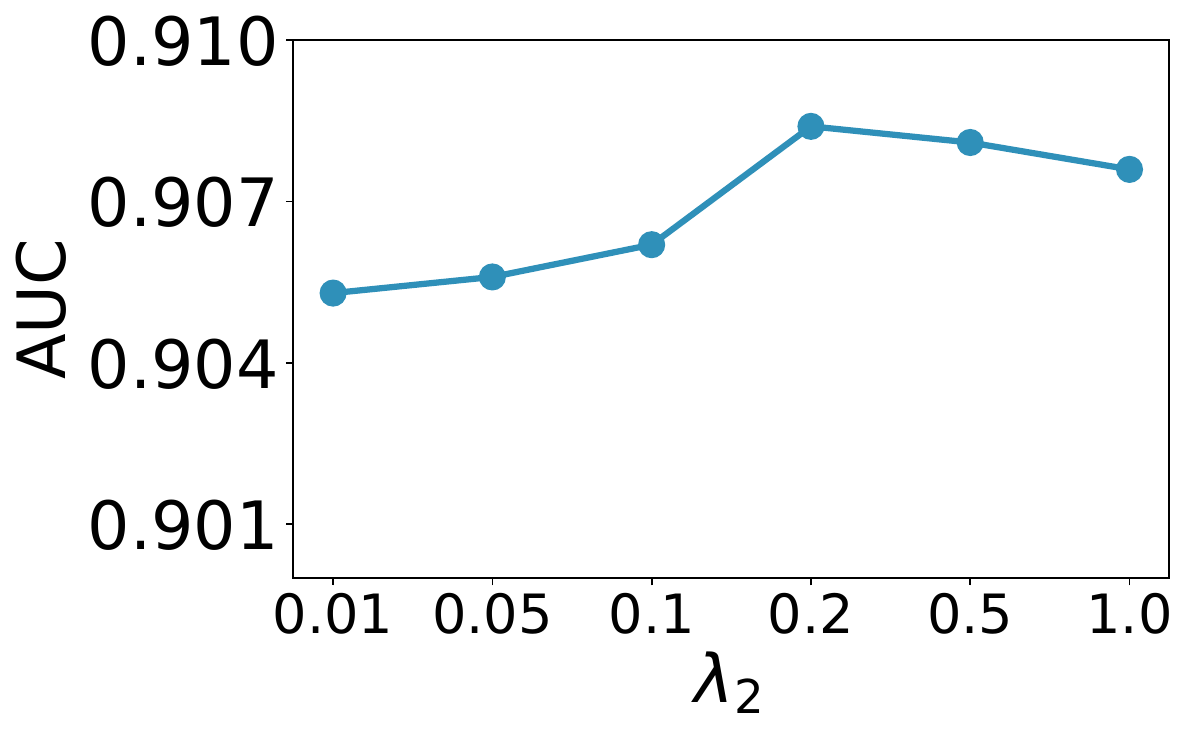}\label{fig:lambda2}
    }
    \vspace{-1.em}
    \caption{Result variation with different parameters.}
    \label{fig:var}
    \vspace{-1.em}
\end{figure}

From the variation trends of Figure~\ref{fig:var}, we could observe that: (1) The better results are achieved when setting $\lambda_1$ and $\tau$ to a suitable value. Too large (e.g., 1.0) or small (e.g., 0.01) value will lead to a performance drop.
(2) The neighbor-target mutual learning can benefit the model performance and $\lambda_2 = 0.2$ generally yields the best result. It should be noted that for each hyper-parameter, a tuning process is conducted while ensuring that the remaining hyper-parameters are set to their optimal values.

\subsection{Online A/B Testing}\label{sec:ab}
We deploy our proposed model on the Alipay search platform providing search service of mini apps, and conduct
online A/B testing by replacing the in-house version of TextGNN with our proposal. Both experiments take about 3\% proportion of Alipay search traffic for two weeks. As a result, the proposed model improves valid PV-CTR\footnote{the number of valid clicks divided by the number of searches} by 0.72\% on average and decreases the online serving latency from 23ms to 13ms approximately.
And the results of human annotations show the model can reduce the rate of irrelevant results by 0.66\% points on average. The results demonstrate that our proposal can improve the experience of users with low latency.

\section{Conclusion}\label{sec:con}
This paper studies the relevance modeling problem by combining textual information and behavior data for achieving promising performance. The novel model BARL-ASe is proposed, which innovatively develops the multi-level co-attention to learn fine-grained information and introduces neighbor-target self-supervised learning for improving representations. For dealing with long-tail query-item pairs, we explore an industrial application through the combination of neighbor-dependent/independent models denoted as BARL-ASe$_+$. The comprehensive experiments on real-world industry data and online A/B testing validate the superiority of our proposal and the effectiveness of its main components.

% \begin{acks}
% \end{acks}

\bibliographystyle{ACM-Reference-Format}
\balance
\bibliography{references}

\end{document}